\documentclass[twocolumn,pra,superscriptaddress]{revtex4}
\pdfoutput=1
\usepackage{amsmath}
\usepackage{graphicx}
\usepackage{color}
\usepackage{graphicx}
\usepackage{amssymb}
\usepackage{bm}
\usepackage{subfigure}
\newcommand{\hide}[1]{}

\newcommand{\eq}[1]{Eq.\,(\ref{#1})}

\newcommand{\fig}[1]{Fig.\,\ref{#1}}

\newcommand{\ket}[1]{\ensuremath{\left| #1 \right\rangle}}
\newcommand{\bra}[1]{\ensuremath{\left\langle #1 \right|}}

\newcommand{\matrixelem}[3]{\ensuremath{\left\langle #1 \right|#2\left|#3\right\rangle}}

\begin{document}

\title{Single photon nonlinearities using arrays of cold polar molecules}
\author{R. M. Rajapakse}
\affiliation{Department of Physics, University of Connecticut,
Storrs, CT 06269}
\author{T. Bragdon}
\affiliation{Department of Physics, University of Connecticut,
Storrs, CT 06269}
\author{A. M. Rey}
 \affiliation{JILA, National Institute of Standards and Technology and University of Colorado, Boulder, CO 80309-0440}
\author{T. Calarco}
 \affiliation{Institute for Quantum Information Processing, University of Ulm, D-89069 Ulm, Germany}
\author{S. F. Yelin}
\affiliation{Department of Physics, University of Connecticut,
Storrs, CT 06269} \affiliation{ITAMP, Harvard-Smithsonian Center
for Astrophysics, Cambridge, MA 02138}

\date{\today}

\begin{abstract}
We model single photon nonlinearities resulting from the dipole-dipole interactions of cold polar molecules. We propose utilizing ``dark state polaritons'' to effectively couple photon and molecular states; through this framework, coherent control of the nonlinearity can be expressed and potentially used in an optical quantum computation architecture.  Due to the dipole-dipole interaction the photons pick up a measurable nonlinear phase even in the single photon regime. A manifold of protected symmetric eigenstates is used as basis.  Depending on the implementation, major sources of decoherence result from non-symmetric interactions and phonon dispersion. We discuss the strength of the nonlinearity per photon and the feasibility of this system.
\end{abstract}

\maketitle
\section{Introduction}
%
Coherent control of optical nonlinearities at the single photon level is a burgeoning topic in quantum optics research. Utilizing state-preserving techniques, it is suggested that one can implement two-qubit quantum logic gates in a feasibly robust optical quantum computational framework \cite{Lukin,kurizki}.

Cold polar molecules are excellent candidates as a mediating medium due to their field-dependent intermolecular interaction properties \cite{Demille,Lee,Jaksch,Andre,YelinKirby}. They have been suggested for quantum computation architectures since they embody advantages of both neutral atoms and trapped ions, viz. long coherence times and strong interactions, respectively.


Advances in preparation (cooling and trapping) of molecular ensembles in their electronic, vibrational, and rotational ground states \cite{Jin} would allow for single state manipulation in a characteristically rich level structure.  Notably, recent work by B\"uchler et al. \cite{Buchler} predicts novel, controllable superfluid and crystalline phase transitions from dipolar gases. The latter could suppress dephasing from short range collisions in high density traps. The anisotropic and long-range form of the dipole-dipole interaction is responsible for the bulk of advances in controlling molecular samples \cite{Jaksch}.

In this paper, we investigate cold polar molecular gases in one- and two-dimensional arrays.  We describe single-photon nonlinearities resulting from the intermolecular dipole-dipole interaction. We apply ``slow'' and ``stored'' light methodology for coherent state transfer. Next we calculate the resultant nonlinear phase in the context of collective excitations in an optically thick media.

Here we are primarily concerned with exploring feasibility of coherent control over the resultant nonlinear phase evolution of intermolecular dipole-dipole interactions. Familiar implementations for the system under discussion include stripline cavities, optical lattices, Wigner crystals, hollow fibers, or molecules on surfaces. Then, we investigate the most significant decoherence effects for implementation in a trap architecture or in a
crystalline phase.

\section{Single-photon nonlinearity}
Photons do not interact. However, effective interaction can be achieved by utilizing state-preserving light-matter couplings to nonlinear media, wherein matter-matter interactions effectuate photon-photon interactions without destroying the state information of the incident coherent fields.

The proposed mechanism is as follows: Photons are efficiently
and coherently coupled into the molecular medium in the form of
``slow-light polaritons.'' The molecule part of these
light-molecule coupled excitations is ``switched'' from the
zero-dipole rotational ground state into a high-dipole rotational
superposition state. The resulting dipole-dipole interaction therefore adds
a nonlinear phase to the polaritons that the photons retain on exiting
the medium. We note that the nonlinear phase is thus proportional to
the interaction time inside the medium and thus to the propagation time of the polaritons. Therefore control over the phase
is exercised by manipulating the propagation velocity of the light in the medium.

\begin{figure}[h]
\centerline{\includegraphics[width=0.7\linewidth]{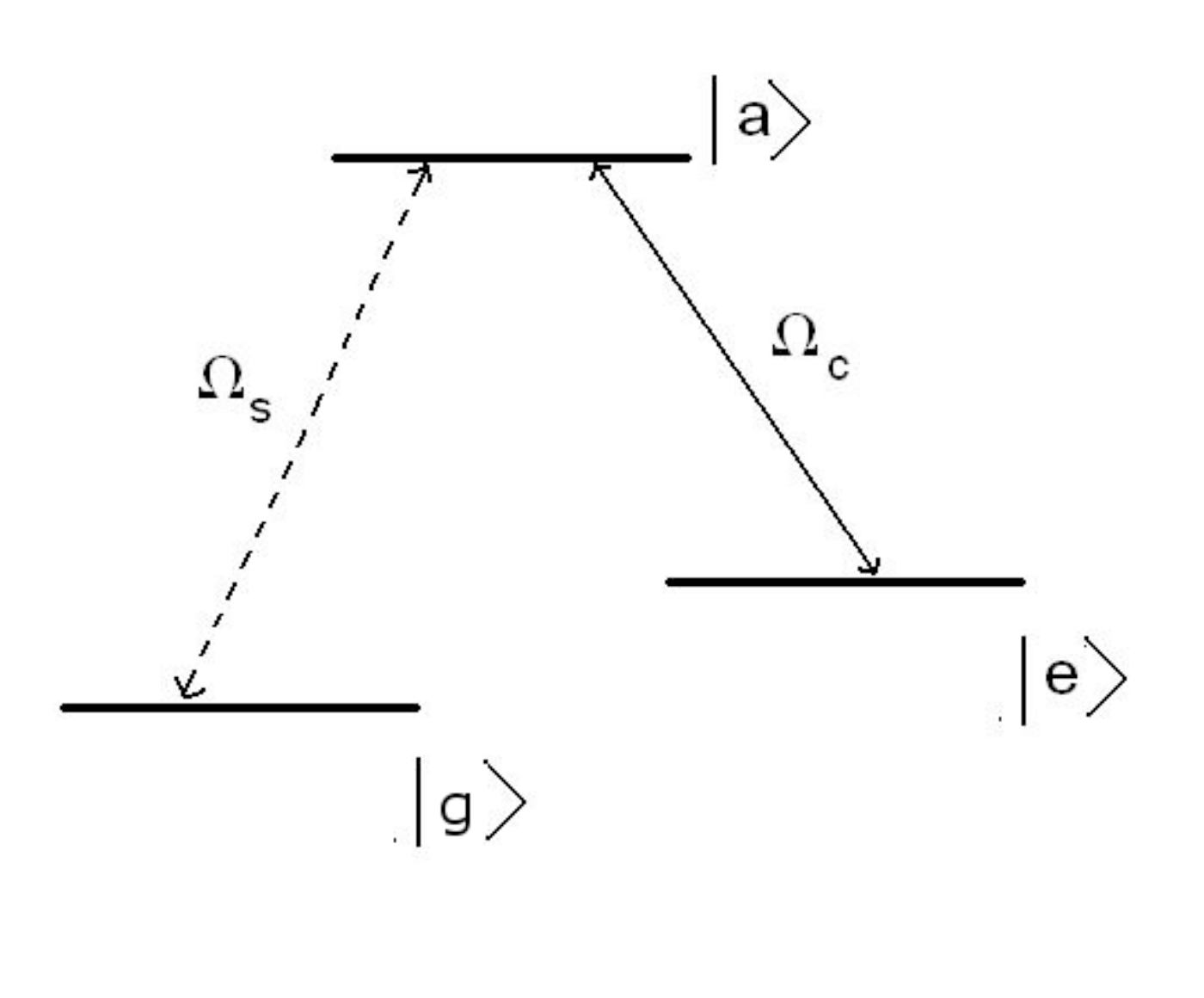}}
\caption{ Level scheme for utilizing slow-light polaritons. \ket{e} and \ket{a} can be coupled via a two-photon transition, or \ket{e} can be a mixed parity state.}\label{threelev}
\end{figure}

Electromagnetically induced transparency (EIT) -based slow light polaritons \cite{Slow,lukinrev} are collective states of
matter-light superposition that can be achieved by using a $\Lambda$-type system.
Polaritons are the coupled exchanges of the signal field $\Omega_s$
and the superposed $\ket{g}$ and $\ket{e}$ ground states (see
\fig{threelev}). $\ket{g}$ and $\ket{e}$ would typically be the $\ket{J, M_J}$ rotational states of the ground state molecules, where $M_J$ is the projection of $J$ on the $z$-axis. The coupling field $\Omega_c$ controls the slow group velocity of the polaritons. 

The interacting states $\ket{g}$ and $\ket{e}$ in our system are neighboring rotational levels of dipolar molecules.
Since dipolar interactions exchange virtual photons, interacting states must have opposite parity.  This can be accomplished
via an expanded $\Lambda$-type system with a strong Raman transition
involving a two-photon transition as $\Omega_c$, or alternatively the
use of mixed-parity states for the excited state $\ket{a}$. For simplicity we refer to
the whole molecular system as an effective two-level system,
consisting of $\ket{e}$ and $\ket{g}$, as is usually done in the
context of slow-light polaritons \cite{Slow}. We adopt a natural
shorthand, $\ket{g_i}$ and $\ket{e_i}$ respectively for the ground and excited
state of the $i$th molecule.

It is convenient to introduce
 collective states  denoted as $|j,m \rangle$. These states are eigenstates of the collective operators $\hat{J}^{2}$ and $\hat{J}_{z}$, where ${\hat{J}} _{\alpha
}=\frac{1}{2}\sum _i\hat{\sigma}_{i}^{\alpha }$,  $\alpha =x,y,z$ and $\hat{\sigma}_{i}^{\alpha }$ are   Pauli
 operators acting on the   $i^{th}$ molecule: $\hat{\sigma}_{i}^{x }=\ket{g_i}\bra{e_i}+\ket{e_i}\bra{g_i}$,
 $\hat{\sigma}_{i}^{y }=i(\ket{g_i}\bra{e_i}-\ket{e_i}\bra{g_i})$, $\hat{\sigma}_{i}^{z }=\ket{e_i}\bra{e_i}-\ket{g_i}\bra{g_i}$.   $\ket{j,m}$ states satisfy  the eigenvalue relations $
\hat{J}^{2}\ket{j,m}=j(j+1)\ket{j,m} $ and $
{\hat{J}^{}}_{z}\ket{j,m}=m\ket{j,m}$, with $
j=N/2,\dots ,0$ and $-j\leq m\leq j$.

Among these states we are particular interested in the fully symmetric Dicke-like states which  lie on the surface  of  the  Bloch sphere with maximal radius $j=N/2$  and   are  totally  symmetric, i.e.
invariant with respect to particle  permutations. We denote them as $|n\rangle =|N/2,-N/2+n \rangle $
to emphasize that they  correspond to  $n$-photon  collective excitation. The corresponding $n=0,1,2$ are explicitly  given by

\begin{eqnarray}
\ket{0}&=&\ket{g_{1},\ldots,g_{N}}\nonumber \\
\ket{1}&=&\frac{1}{\sqrt{N}}\sum_{i=1}^{N}\ket{g_{1},\ldots,e_{i},\ldots,g_{N}}\\
\ket{2}&=&\sqrt{\frac{\scriptstyle 2}{\scriptstyle N(N-1)}}\sum_{i< j}\ket{g_{1},\ldots,e_{i},\ldots,e_{j},\ldots,g_{N}}\nonumber
\label{gestates}
\end{eqnarray}
In our case, cold polar molecular gases e.g., SrO \cite{Demille} or CaF \cite{MicheliZoller} constitute the nonlinear medium. The nonlinearity is expressed through dipole-dipole interactions. Imagine first the ideal case when such interactions between molecules generate an effective Hamiltonian of the type
\begin{eqnarray}
\hat{V}_{dd}= \chi  \hat{J}^{2}_{z}. \label{jz}
\end{eqnarray} Since the states  $|n\rangle$ are  eigenstates of $\hat{V}_{dd}$, the collective dynamics can be fully accounted for  by their phase evolution $\theta_{n}$:
 \begin{equation}
 \hbar \theta_{n}(t)=\matrixelem{n}{\hat{V}_{dd}t}{n}=\chi (N/2-n)^2 t
\label{thetan}
\end{equation}In general to  characterize the medium nonlinearity we would have to include photon coupling states with \(n>2\), but in the scheme of optical quantum computation it  is sufficient to implement two-qubit controlled phase operation\ \textendash\ provided appropriate single qubit gates. The latter can be realized if the accumulated non-linear phase $\Theta(t)$ acquired by the polaritons equals $\pi$ at the time when they exit the medium. $\Theta(t)$ is the difference between the phase picked up by two concurrent excitations and the sum of the phases that each individual excitation would independently pick up in the absence of the other. It is defined as
\begin{equation}
\Theta(t)=(\theta_2(t)-\theta_0(t))-2(\theta_1(t)-\theta_0(t))=\theta_{2}-2\theta_{1}+\theta_{0}.
\label{phiNL}
\end{equation} In other words, $\Theta(t)$ quantifies the departure from a linear regime, that is, one in which the interaction between the two excitations is absent and therefore their individual phases simply sum up. From \eq{jz} the latter condition is satisfied if  $2 \chi t_\pi =\hbar \pi$.
Therefore in this ideal case  establishing a deterministic controlled phase operation only  requires  coherent control of the propagation and/or storage time of the polariton in the dipolar medium. This corresponds to manipulating the control fields that establish the conditions for the ``slow light" propagation.

However, dipolar interactions do not generate a Hamiltonian of the type described by \eq{jz} and instead  the dipole-dipole interaction is given by
\begin{equation}
\hat{V}_{dd}^{(1D)}=\frac{1}{8\pi \epsilon_0}\sum_{i\neq j}\frac{\hat{\bm{\mu}}_i\cdot\hat{\bm{\mu}}_i - 3(\hat{\bm{\mu}}_i\cdot \bm{r}_j^0)(\hat{\bm{\mu}}_j\cdot \bm{r}_i^0)}{|\bm{r}_i^0-\bm{r}_j^0|^{3}}
\label{dip1}
\end{equation} with  $\bm{\mu}_i$ the dipole moment of the molecule at site $\bm{r}_i^0$. Here   we have assumed that the molecules are at fixed positions determined for example by a superimposed  external optical lattice potential.

For the simplest situation when  both $\ket{g}$ and $\ket{e}$ states are pure rotational states and  have zero dipole moment \(\bm{\mu_{gg}}=\bm{\mu_{ee}}=0\),  the dipole-dipole interaction is governed by the $\ket{g}\leftrightarrow\ket{e}$ transition dipole moment which can be formally written as:
\begin{eqnarray}
\hat{\bm{\mu}}_i&=&\bm{\mu_{ge}}\ket{g_i}\bra{e_i}+\bm{\mu_{eg}}\ket{e_i}\bra{g_i}\nonumber \\
&\equiv& \bm{\mu_{ge}}\hat{\sigma}_{i}^{-}+\bm{\mu_{eg}}\hat{\sigma}_{i}^{+},
\label{dipmoment}
\end{eqnarray}
with $\bm{\mu_{ab}} \equiv \mu_0 \bra{a} \bm{e_r} \ket{b}$.  Assuming the interacting dipoles are aligned in parallel, which is possible in 1D and 2D geometries, this leads to $(\hat{\bm{\mu}}_i\cdot \bm{r}_j^0)=0$, and neglecting counter-rotating terms \(\hat{\sigma}_{i}^{+}\hat{\sigma}_{i}^{+}\), \(\hat{\sigma}_{i}^{-}\hat{\sigma}_{i}^{-}\)  the  interaction potential becomes
\begin{equation}
 \hat{V}_{dd} =\frac{|\mu_{eg}|^{2}}{8\pi \epsilon_0} \sum_{i\neq  j}\frac{\hat{\sigma}_{i}^{+}\hat{\sigma}_{j}^{-}+\hat{\sigma}_{i}^{-}\hat{\sigma}_{j}^{+}}{|\bm{r}_i^0-\bm{r}_j^0|^{3}}.
\label{Vdd}
\end{equation}

$\hat{V}_{dd}$ is not $SU(2)$ symmetric and consequently the collective Dicke states  are  not eigenstates of it.
Exceptions are $|0\rangle$ and $|1\rangle$ states which do remain eigestates of  $\hat{V}_{dd}$.
This implies that the dynamical evolution of  $\ket{n}$ for $n\ge 2$ not only acquires a time-dependent phase, but in addition transitions to other states
 outside $j=N/2$ will take place.
These transitions  will affect the implementation of the
phase gate which relies on remaining on the Dicke manifold.

Ignoring for the moment the ''leakage'' outside the Dicke states and focussing  only the projection $\mathcal {P}$ of $\hat{V}_{dd}$ on the Dicke manifold, which is given by \cite{anamaria}
\begin{eqnarray}
\mathcal {P}\hat{V}_{dd}&=&\chi_{\rm eff}  \hat{J}^{2}_{z}+ const\\
\chi_{\rm eff} &=& \frac{2\kappa}{ N(N-1)} \sum_{i\neq  j}\frac{a^3}{|\bm{r}_i^0-\bm{r}_j^0|^{3}} \label{eff}\\
\text{where} && \kappa = \frac{|\mu_{eg}|^{2}}{8\pi a^3 \epsilon_0}\nonumber
\end{eqnarray}
and $a$ the lattice constant   of the molecular array.
Now one can estimate the propagation time required for implementing the phase gate as $t_\pi=\hbar \pi/ (2 \chi_{\rm eff})$. The resulting expression gives $\chi_{\rm eff}\propto\,^1\!/_N$. It clearly shows that there is an optimization to undertake regarding the number of molecules in the array: On one hand, there must be enough molecules to create sufficient optical depth to couple-in the polaritons \cite{vladan}. On the other hand, in order to maximize the nonlinearity in \eq{eq:nonlin1D}, less molecules are better.

In a one dimensional (1D) molecular array, one can  analytically evaluate the nonlinear phase
factor $\Theta$ from \eq{eff}.   It is given by :
\begin{equation}
\Theta^{(1D)}\approx\frac{4\,\kappa \,t\, \zeta[3]}{\hbar(N-1)}
\label{eq:nonlin1D}
\end{equation}
where \(\zeta[3]=\lim_{N\to\infty}\sum_{i=1}^{i=N-1}i^{-3}\approx 1.2\).

Given aforementioned assumptions
about dipole alignment, for the two-dimensional square lattice \footnote{For simplicity
we assumed a square lattice. This
assumption does not change the overall picture of the dynamics if one looks at other types of 2D lattices, such as triangular ones.} only a change of lattice vectors \(\bm{r}_i^0=y_i \bm{e_y}+z_i \bm{e_z}\) is required. In this case and assuming  periodic boundary conditions one obtains
\begin{equation}
 \Theta^{(2D)} \approx 2 \Theta^{(1D)}
\end{equation}

\begin{figure}
\centering
\subfigure[]{\includegraphics[width=0.8\linewidth]{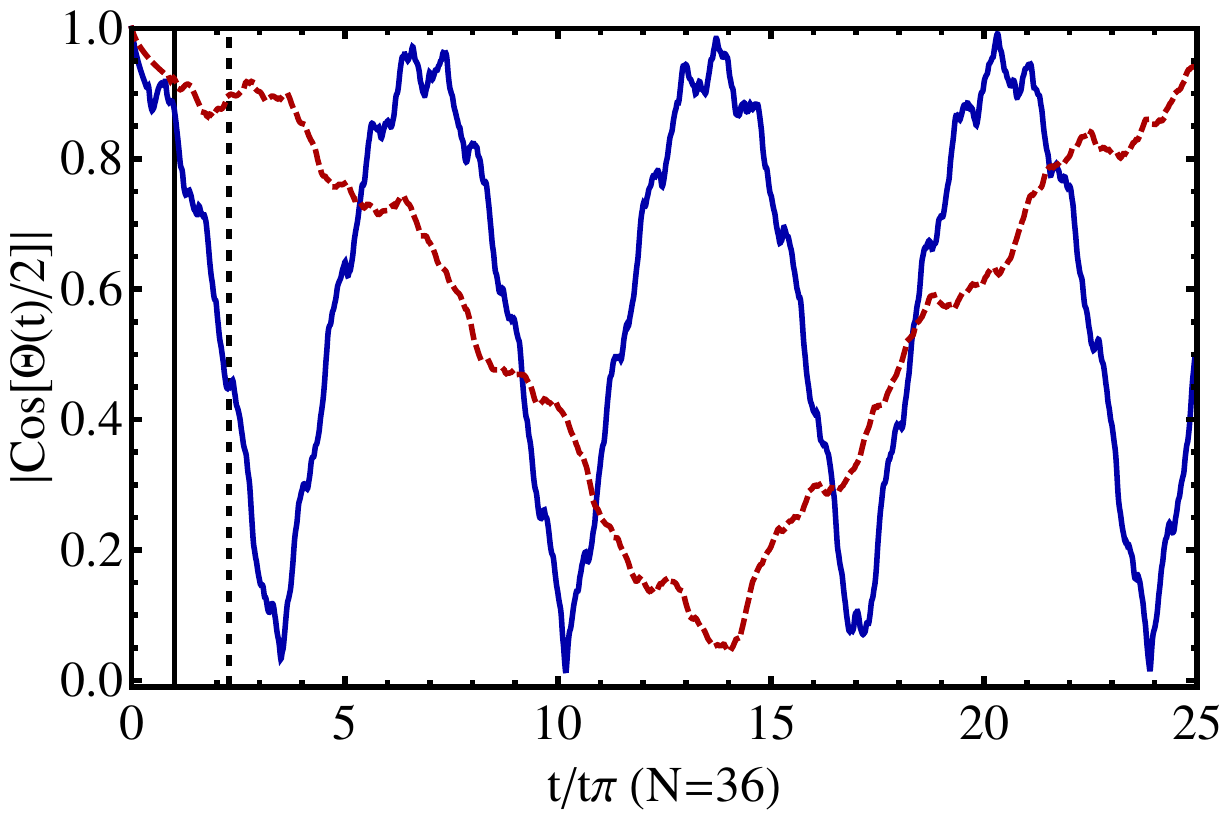}}\quad
\subfigure[]{\includegraphics[width=0.8\linewidth]{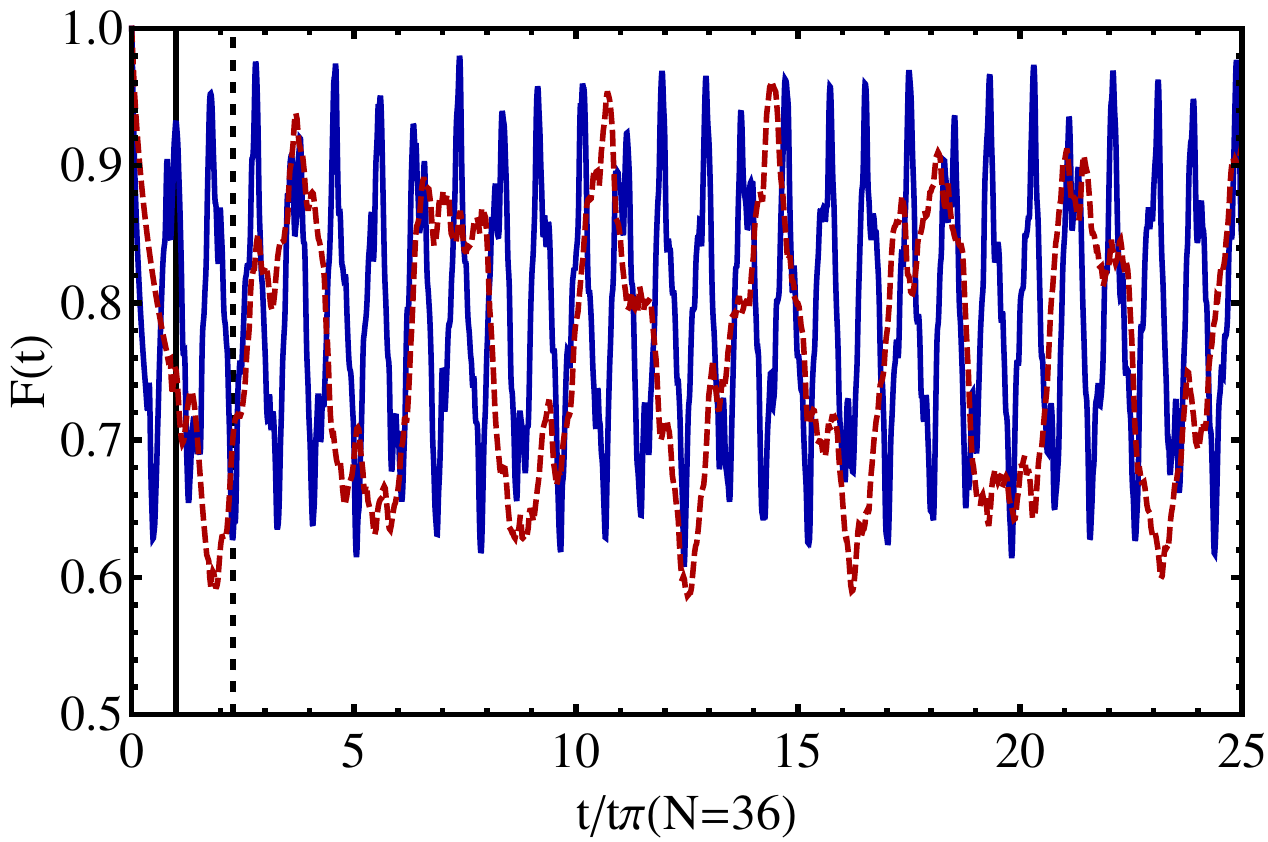}}
\caption{(Color Online): (a) Nonlinear phase  as a function of time for a 1D array with $N=36$ molecules (blue solid line) and $N=81$ (red dashed line).   Here $t_{\pi}(N=36)=\hbar \pi/ (2 \chi_{\rm eff})$  is the expected phase gate time  for the $N=36$ system  calculated by projecting the dipole Hamiltonian (\eq{dip1}) onto the Dicke manifold. It is obvious from this graph, that the projected $\Theta=\pi$--phase time ($t=t_\pi$, solid black line for $N=36$, broken for $N=81$) deviates strongly from the exact one ($t\approx 3.5\,t_\pi$).
The main reason for this deviation is the importance of transition processes {\it out of the Dicke manifold} as confirmed in panel (b) where we plot
the fidelity to stay in the state $\ket{2}$.}  \label{decayprob}
\end{figure}

\section{Decoherence}

\subsection{Decay Out of Symmetric manifolds}
\label{sec:important}

Decay out of symmetric manifolds and phonon-like effects in the  Wigner crystal implementations can be significant. We will discuss symmetric manifolds at present, leaving the phonon-like decoherence effects to a later section.

As mentioned in previous session, Dicke states are a good basis only if the relevant Hamiltonian is spherically symmetric (SU(2) symmetric). This is not the case  for  $\hat V_{dd}$,  and in particular the state $|2\rangle$ will decay during the time evolution inducing  decoherence.

To estimate the  decay probability from an initial Dicke eigenstate during the dynamical evolution we calculate  the fidelity $F(t)$
\begin{equation}
F(t)=\frac{ {\mid \langle{\psi(t)} \ket{2}\mid}^2}{ {\mid \langle{\psi(0)} \ket{2}\mid}^2}.
\label{Ft}
\end{equation} where $\dot{\ket{\psi(t)}}=-\frac{i}{\hbar}\hat V_{dd}\ket{\psi(t)}$, is the  time evolving state under $\hat V_{dd}$ and  $\ket{\psi(0)}=C_2(0)\ket{2}+ C_1(0)\ket{1}+ C_0(0)\ket{0}$. This quantity is plotted for two different 1D sample sizes  in \fig{decayprob}. In the same figure we also show the non-linear phase $\Theta$ accumulated by the evolving state. We numerically evaluated it as

\begin{equation}
\cos[\Theta(t)/2]=\frac{C_0^*(t) C_2(t)+(C_0^*(t) C_1(t))^2}{2}.
\label{thet}
\end{equation}where $C_n(t)=(\langle{\psi(t)} \ket{n})/\langle{\psi(0)} \ket{n}$ are the  projections of  the evolving state into the corresponding Dicke states. The relevance of decoherence effects and the departure of the pure phase accumulation can be clearly observed in \fig{decayprob}. The plot shows not only a distorted evolution of the non-linear phase but also a different time dynamics since  the non-linear  phase approaches $\pi$ at a time very different from the expected $t_\pi$ (See Eq. (10)) .

One could effectively remove the mixture of $j$ manifolds and improve the fidelity of the phase gate by the addition of an external electric field. This procedure, which we will describe in the following section, generally establishes a Many-body Protected Manifold (MPM) \cite{anamaria} which helps to eliminate or mitigate  decoherence effects.


\subsection{Phonon-like effects}
 In the presence of an external DC field, which induce  repulsive dipole-dipole interactions in the ground and excited states, (\(\mu_{gg},\mu_{ee}\neq0\)),  molecules can assemble themselves in a  Wigner crystal.
Attractive interactions along the remaining directions can be  suppressed by a strong transverse confinement  \cite{Buchler}. In a Wigner crystal implementation, there exists another notable decoherence effect, which can be analyzed using a phonon formalism, cf.~\cite{rabl2}. In a realistic crystalline phase, the molecules are not fixed frozen. Phonon-like effects will add to the decay described in the previous section as their energy provides a coupling between the symmetric and non-symmetric states.  Details of decoherence due to phonons will also be treated in subsequent sections.

\subsection{Finite pulse effects}
In the previous analysis we have assumed that  Dicke states are the result of  the coherent light-molecule interactions. However finite pulse effects can introduce inhomogeneity and can  lead to non-zero initial population of states out-side the Dicke manifold. The latter will cause additional  decoherence and will  degrade  the phase gate. The corrections can be quantitatively understood by noticing that while slow light polaritons have linear dispersion in a linear medium,  the nonlinear interaction  adds a  dispersion relation:
\begin{equation}
E(\bm {k })=\hbar \omega_{\bm {k }}=\kappa
\sum_{j} \frac{4}{|\bm{r}_0^0-\bm{r}_j^0|^{3}} \sin^2(\frac{1}{2}{\bm {k \cdot r}}_j^0). \label{spinw}
\end{equation}
For 1D, \eq{spinw} can be rewritten as
\begin{equation}
\hbar\omega_{k\rightarrow 0}^{1D}\rightarrow \kappa\left(\left(-3+ 2\ln ka\right) (ka)^2 + O(k)^3\right)
\label{Ek1}\end{equation}The non-linear terms in $k$ present in  the 1D dispersion relation  will degrade the phase gate. They however can  be mitigated by using long pulses or a (ring) cavity, where $k=0$.

 In 2D on the contrary the low energy excitations  scale as
\begin{equation}
\hbar\omega_{k\rightarrow 0}^{2D} \rightarrow 3.27 \kappa |\mathbf{k}a|
\label{Ek2}\end{equation} showing that in the 2D case, at least in the long-wave limit,   the spectrum remains linear  and thus decoherence due to finite pulse effects becomes less important.

\section{Manybody Protected Manifold (MPM)}

As the next step, we include a tunable DC electric field which thereby induces a dipole moment in the ground and excited states of the molecule ($\mu_{gg},\mu_{ee} \neq 0$). This effect then augments the dipole-dipole interaction among our collective Dicke-like states in a way that enables perturbative treatment of the non-spherically symmetric part of the interaction, thus reinstating $\ket{j,m}$ as good eigenstates for the system:
\begin{eqnarray}
\hat{V} &=& H_H+ H_I\;=\\
&=&\frac{\kappa}{2} \sum_{i\neq j} \frac{a^3}{\left|\bm{r}_i^0-\bm{r}_j^0\right|^3}\hat{\bm{\sigma}}_i \cdot \hat{\bm{\sigma}}_j-\frac{\xi}{2} \sum_{i\neq j} \frac{a^3}{\left|\bm{r}_i^0-\bm{r}_j^0\right|^3}\hat{\sigma}_i^z \hat{\sigma}_j^z,\nonumber\\
\text{where }\hspace{-5mm}&& \xi= \frac{|\mu_{eg}|^2-\frac{1}{2}(\mu_{ee}-\mu_{gg})^2}{8 \pi a^3 \epsilon_0}.\nonumber
\label{VddDC}
\end{eqnarray}
Here $H_H$ is the spherically symmetric (Heisenberg) part of the Hamiltonian $V$, and $H_I$ is the non-symmetric (Ising) part.

 If at  $t=0$ an  initial
state is prepared within the $j=N/2$   manifold, a  perturbative analysis predicts
 that for times $t$ such that $\kappa t/\hbar <\xi/\kappa$,
 $\hat{H}_{H}$ confines the dynamics to the Dicke  manifold  and
transitions outside it   can be neglected. In other words the Dicke manifold becomes protected by the many-body interactions and  only
the projection of  $\hat{H}_{I}$  on it, which corresponds   to
 $\mathcal {P}\hat{H}_{I}=- {\tilde{\chi} }_{\rm eff} \hat{J}^{2}_{z}+ {\rm{const}}$, with

 $\tilde\chi_{\rm eff} \,=\, \frac{\xi}{\kappa}\chi_{\rm eff}$
 becomes effective.  As a consequence $H_I$
acts as the desired ideal ``phase gate'' Hamiltonian.

\begin{figure}[h]
\centerline {\includegraphics[width=\linewidth]{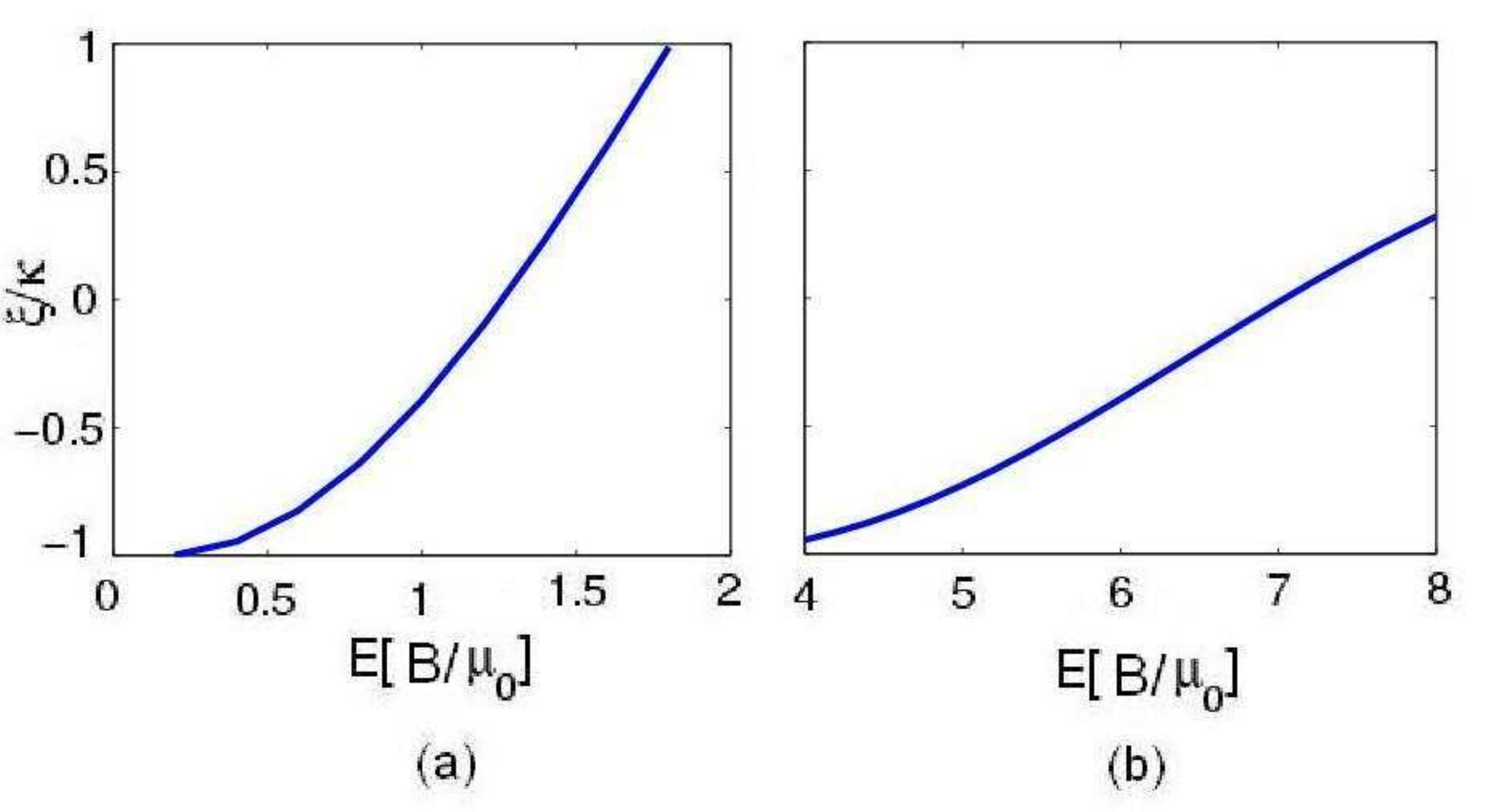}}
\caption{$\xi / \kappa$ for varying $E [B / \mu_0]$ in SrO. (a) $\ket{g_i}=\ket{J_{\rm SrO},M_{J,{\rm SrO}}}=\ket{0,0}_i$ and $\ket{e_i}=\ket{1,0}_i$. (b) $\ket{g_i}=\ket{1,0}_i$ and $\ket{e_i}=\ket{2,0}_i$. $B$ is the rotational constant of the molecule, and $\mu_0$ is the maximum ground state dipole-moment, in our case assumed to be $\mu_0=|\bm{\mu}_{eg}|$.}
\label{xi_kappa}
\end{figure}

The relative strength of the $H_H$ and $H_I$ parts of $\hat{V}$ can be manipulated to find values of \(\xi/\kappa\) such that MPM protection is maximized. For example, SrO has a \(^1\Sigma\) ground state with a magnetic moment of 8.89D. Upon selecting ground and excited rotational states with opposing parity, the appropriate values of $\xi$ and $\kappa$ are then obtained by diagonalizing the Stark Hamiltonian for variable electric fields \cite{rabl2}. In \fig{xi_kappa}, we show the ratio of \(\xi / \kappa\) for varying DC field
strength between two rotational levels of SrO. For induced dipole transitions, the biasing electric DC field $E$ depends on both the rotational constant B and the ground state dipole moment $\mu_0$.

The application of an electric field changes the original bare states to dressed states $\ket{g}$ and $\ket{e}$ which are linear superpositions of the bare states. It has to be noted that the addition of a DC field, leading to MPM protection, causes the nonlinearity to be reduced by a factor of $\sim\left|\xi/\kappa\right|$.

\subsection{Decay Out of Manybody Protected manifold into other
manifolds}

In this section we study what occurs if at time $t=0$ we
prepare the system in the $j=N/2$ subspace and let the system
evolve in time in the presence of MPM. This can be written as
\begin{equation}
\ket{\Psi_n(t)} = e^{-\frac{1}{2}\gamma_n(t) t} e^{-i\theta_n (t)} \ket{^N\!/_2 ,-^N\!/_2+n}
\end{equation}
Here we estimate the decay magnitude $\gamma_n$.  Our objectives are to keep $\theta_n$ large while minimizing $\gamma_n$.  As the states with 0 and 1 excitations are eigenstates of the symmetric manifold there is no decay out of the manifolds and therefore $\gamma_0=0$ and $\gamma_1=0$.

Using first-order perturbation theory we can write
\begin{equation}
e^{-\gamma_2 (t)t } \simeq 1 -\frac{1}{\hbar}\sum_{{\bm {k,k'}>0} }\bigl\lvert\int_{0}^{t}d\tau
\mathcal{M}_{{\bm {k,k'}}}e^{i\tau (\omega_{\bm {k}}+\omega_{\bm
k'})}\bigr\rvert^{2}
\label{decay2}
\end{equation}
The quantities
 $\mathcal{M}_{{\bm{ k,k'}}
}=\matrixelem{^N\!/_2,-^N\!/_2+2 }{H_I }{\psi_{\bm {k,k'}>0}}$ are the transition matrix elements to states with $j=N/2-2$ which are the only ones which couple to $|2\rangle$ according to the Wigner-Eckart theorem. To a good approximation  they are given by: $
|\psi_{\bm{k,k'}}\rangle=\frac{1}{\sqrt{N(N-1)}}\sum_{i\neq j} e^{i
({\bm {k \cdot r}^0}_j+ {\bm{k' \cdot r}}_i^0)} \sigma^{-}_j \sigma
^{-}_i|0\rangle$ and  their corresponding excitation energies by $ \hbar(\omega_{\bm {k}}+\omega_{\bm
k'})$, with $\omega_{\bm {k}}$ given by  \eq{spinw}.  $\hbar {\bm k}$, $\hbar {\bm  k'}$ are discrete quasi-momenta, which for a 2D square lattice with lattice spacing $a$ can be written as $ {\bm k}=\frac{2\pi}{a \sqrt{N}}(i,j)$,  $i,j=0,\dots N-1$.
Note that the sum  over  ${\bm k},{\bm k'}$ in $\mathcal{M}_{{\bm{ k,k'}}}$ excludes the state  ${\bm k}={\bm k'}=0$ since $\psi_{\bm {0,0}}$ is just $|2\rangle$.

After some algebra, one can show that
$\mathcal{M}_{\bm{k,k'}} = \dfrac{4\xi}{N} F_{\bm k}\delta_{\bm {k,-k'}}$ where  $F_{\bm {k}}$ is the Fourier series of
$\lvert {\bm r}_i^0-{\bm r}_j^0\rvert^{-3}$, i.e.  $ F_{\bm {k}}= a^3 \sum_{j} \lvert {\bm r}_0^0
-{\bm r}_j^0\rvert^{-3} \cos(\bm {k \cdot r}_j^0)$.  Replacing the latter equation  in \eq{decay2} yields the following expression for the decay rate
\begin{equation}
e^{-\gamma_2( t)t} \simeq  1-\frac{16{\xi}^2}{N^2}  \sum_{\bm {k}>0 }|F_{\bm {k}}|^2\,\frac{\sin^2( \omega_{\bm {k}}
t)}{\hbar^2\omega_{\bm {k}}^2}.\label{decay}
\end{equation}

To get a general idea on the decay rate behavior  we first
use the Fermi- Golden Rule to estimate the decay rate in the
thermodynamic limit, $N\to \infty$ and then  we compare this predictions  with numerical studies for finite size systems. According to the Fermi-Golden rule, at long times  the decay probability evolves linear with  time as  $\gamma_2(t)=\Gamma_2 $:
\begin{eqnarray}
\Gamma_2&\simeq& \frac{ 4\pi \xi^2  }{\hbar N} \int \frac{(a d {\bm{k})^D}|F_{\bm
{k}}|^2}{(2 \pi)^D |\nabla_{\bm {k}} \omega_{\bm {k}}|}\delta(a \bm{k}) ,
\end{eqnarray}The latter relation yields that $\Gamma_2 t_\pi$ diverges in 1D as
\[
\Gamma_2 t_\pi \propto \frac{\xi}{\kappa} \int d(k a) \frac{\delta(ka)}{|ka \log ka|} \to \infty,
\] implying the break down of the Fermi-Golden rule approximation
and emphasizing the issue that in 1D non-symmetric  decoherence effects  are crucial in the large $N$ limit.

In 2D, the situation is  better due to the linear dependence of the
long wave excitations with $k$ and the extra-factor of $k$ in the
density of states. This yields that
\[
\Gamma_2 t_\pi \propto \frac{\xi}{\kappa} \int d(k a) ka \delta (ka) \to 0
\]
and
\begin{equation}
F^{2D}(t_\pi) \to 1  ,
\end{equation}
Consequently as long as $\xi <\kappa$  (which is required for the
validity of our perturbative treatment) and neglecting  other
decoherence effects during the time evolution (which grows linearly
with N) the Fermi-Golden rule  predicts a robust phase gate in 2D.

To validate this predictions we solve the exact many-body dynamics numerically by evolving a system initially prepared in the
Dicke state with $n=2$ under the effective Hamiltonian $\hat V$, and compute the fidelity, $\mathcal{F}$  of remaining
in a Dicke state:
\begin{equation}
\mathcal{F}(t)=\left\lvert \bigl\langle ^N\!/_2,-^N\!/_2+2\big|\Psi(t)\bigr\rangle\right\rvert^2=
e^{-\gamma_2(t)t}.
\label{F1t}
\end{equation}

\begin{figure}
\centering
\subfigure[]{\includegraphics[width=0.9\linewidth]{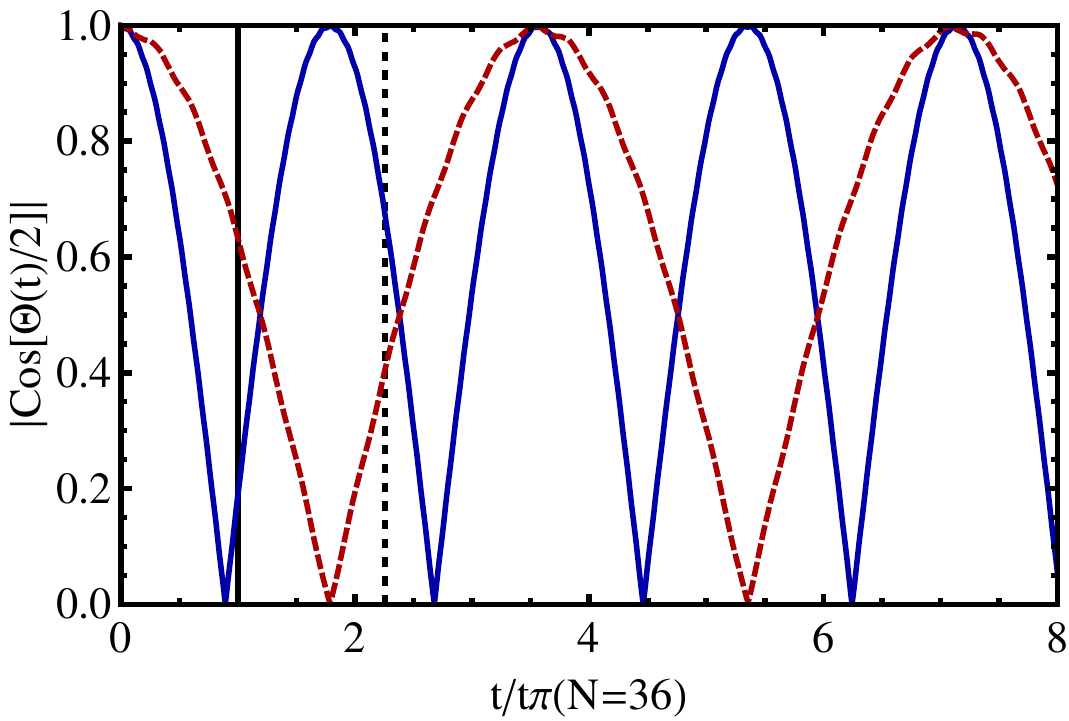}}\quad
\subfigure[]{\includegraphics[width=0.9\linewidth]{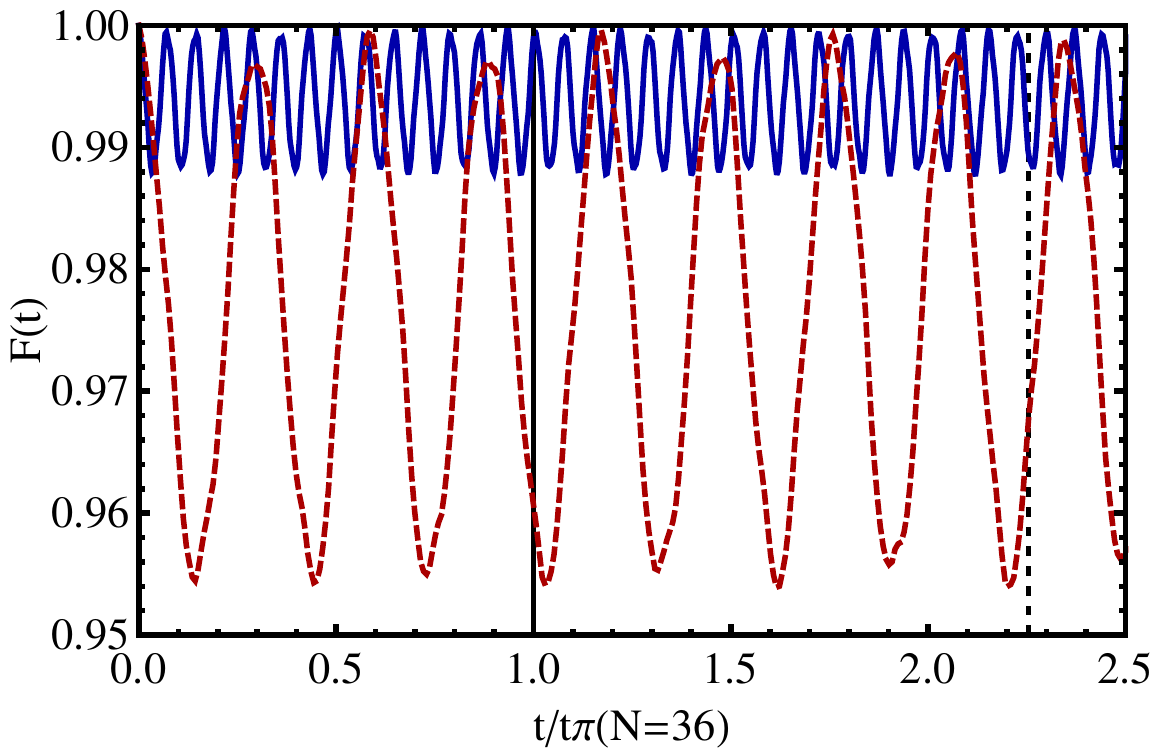}}
\caption{(Color Online): a) Nonlinear phase  as a function of time for a 1D array with $N=36$ molecules (blue solid line) and $N=81$ (red dashed line) in the presence of an external DC field,  $\xi/\kappa=0.05$. The latter is used to implement the MPM.  Here $t_{\pi}(N=36)=\hbar \pi/ (2 \tilde{\chi}_{\rm eff})$  is the expected phase gate time  from our perturbative analysis for the $N=36$ system (indicated by a solid grid line). The corresponding time for the $N=81$ system is indicated by the dashed vertical line. For the two cases the actual time at which the phase gate is accomplished, $|\cos(\Theta/2)|=0$ is close to the calculated $t_{\pi}$ indicating the validity of the perturbative analysis specially for $N=36$. The deviations can be accounted for by higher order corrections in perturbation theory.
The fidelity of remaining in  the $|2\rangle$ state is shown in panel b. The latter decreases as either the ratio $\xi/\kappa$ or $N$ increases, consistently with the Fermi-Golden rule predictions.  }  \label{seven}
\end{figure}

In \fig{seven} we show the 1D dynamics using the
parameters $\xi/\kappa=0.05$ and $N=36$ and $81$ and plot both $\Theta$ and $F(t)$. Note in the presence of the MPM, the time when
the phase gate is implemented  is close to the expected time $t_\pi$. \fig{seven} confirms the prediction that in 1D the fidelity decreases with increasing $N$.  We find
that for moderate $N$ the 1D decay probability increases as
$\approx 0.01\dfrac{\xi^2}{\kappa^2} N^{1.62}$, which is obtained via a fit. This relation
implies that in order to implement a robust gate
\begin{equation}
\frac{\xi^2}{\kappa^2}\ll \frac{100}{N^{1.62}}
\end{equation}
However, by choosing $\xi/\kappa$ small we pay the price of
having slower dynamics and therefore we make the system more
vulnerable to other type of losses.

\begin{figure}
\centering
\subfigure[]{\includegraphics[width=0.9\linewidth]{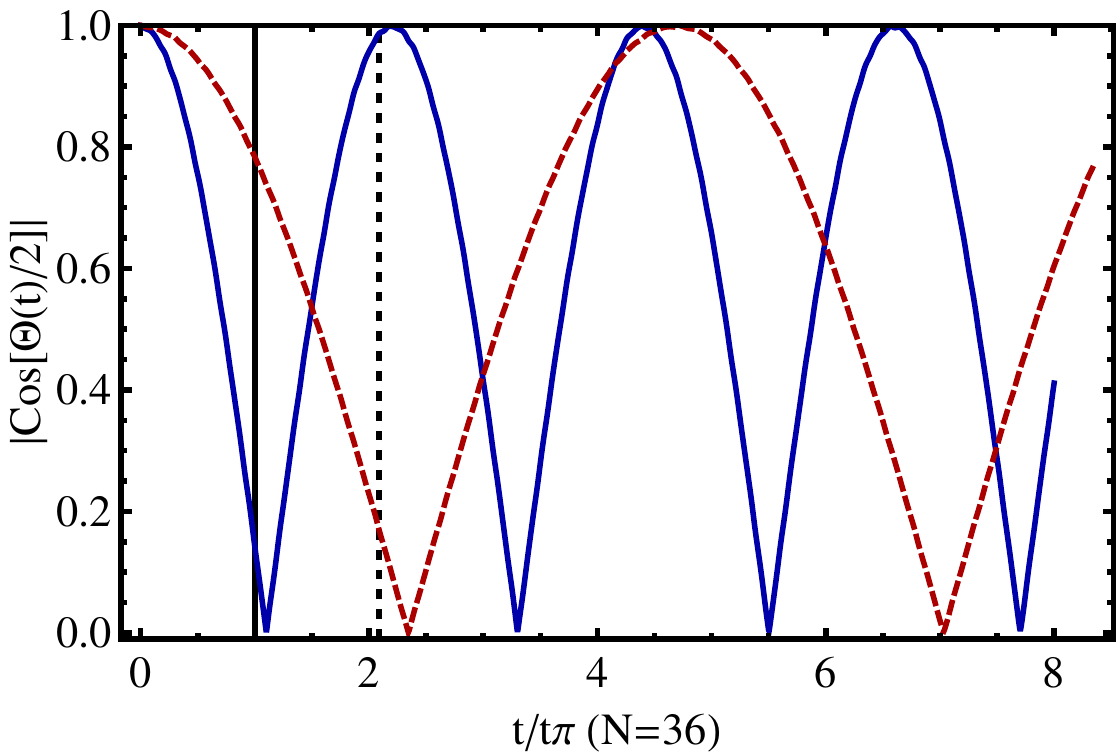}}\quad
\subfigure[]{\includegraphics[width=0.9\linewidth]{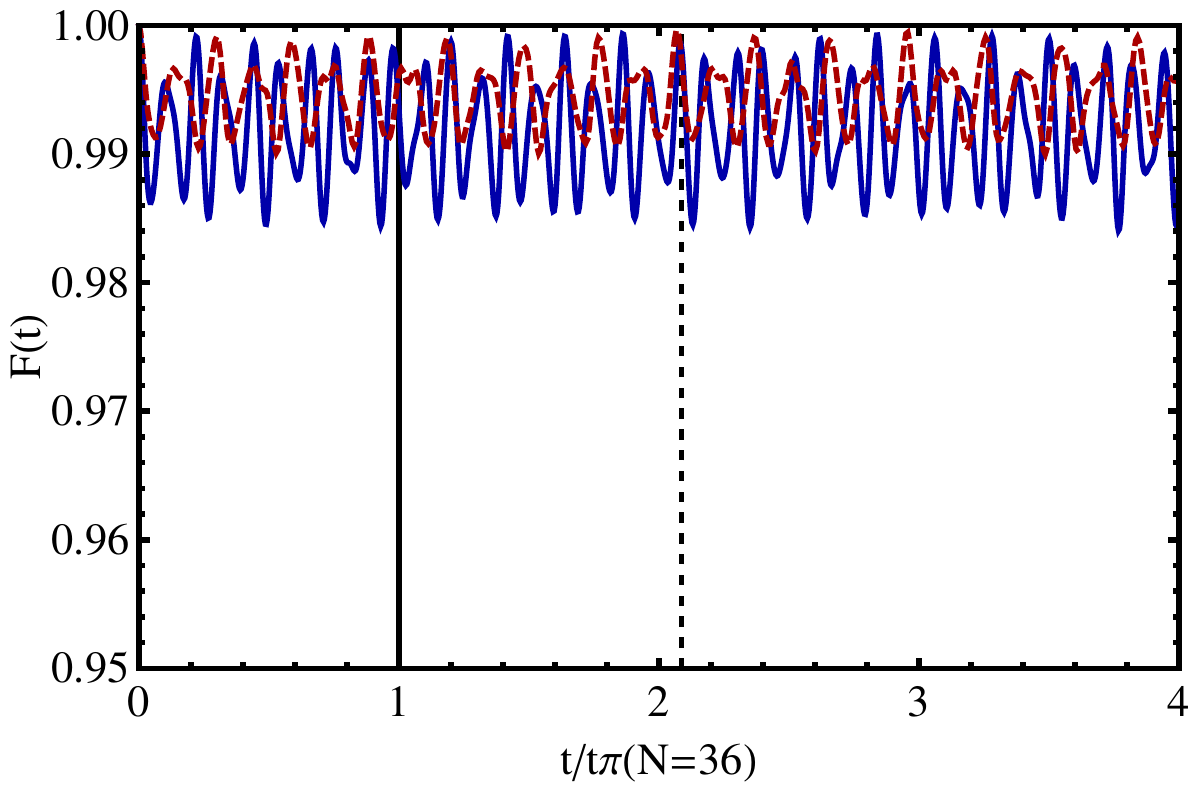}}
\caption{(Color Online): a) Nonlinear phase  as a function of time for a 2D array with $N=36$ molecules (blue solid line) and $N=81$ (red dashed line).  The solid and dashed vertical lines indicated the expected phase gate time according to our perturbative analysis. The ratio of $\xi/\kappa=1$ is outside the perturbative regime however the dynamics  exhibits almost the behavior expected from an ideal $\hat{J}_z^2$ evolution.
The fidelity to remain in  the $|2\rangle$ state is shown in panel b. Note that in 2D the fidelity improves  as  $N$ is increased.  } \label{eight}
\end{figure}

\fig{eight} emphasizes the gain in fidelity obtained by going from 1D to 2D. With  the same number of molecules and even a much larger $\xi/\kappa=1$ the fidelity is much better than in 1D. By numerically evaluating the maximum decay probability we find  it  behaves as $\propto\frac{\xi^2}{\kappa^2} N^{-0.86}$. The decrease in $F$  with increasing $N$  is in agreement with the Fermi-Golden Rule approximation.

\section{Wigner crystal and phonons}
In dense low temperature  systems with sufficiently strong fixed DC electric fields, \cite{Buchler}, one can realize a  self ensemble molecular crystal or  Wigner crystal. In this crystalline phase $|\mu_{gg}|>0$ and  the molecules are localized at
their classical equilibrium positions,  $\bm{r}_i^0$. The latter  form a linear chain
in 1D or a triangular lattice in 2D, with  lattice spacing $a$.
The formation of a Wigner crystal is fundamentally determined by the dimensionless
parameter
\begin{equation}
\beta\,=\,\frac{\rm potential\hspace{1mm}energy}{\rm kinetic \hspace{1mm}energy} \,\equiv\,\frac{U_{dd}}{\hbar^2/( m a^2)},
\end{equation}
 for molecules of mass $m$ for a given density
$\rho = 1/(a)^D$. For $\beta\gg1$ the
dipolar repulsion wins over kinetic energy  stabilizing the crystalline phase.
In contrast to the case where the the localization of the molecules is enforced by an external potential such as an optical lattice,  in the self-assembled
crystalline phase,
molecules are not frozen and they can exhibit collective oscillations (phonons) about their equilibrium positions.
These oscillations can be described
by rewriting the  position operators as $\bm{r_i}=\bm{r}_i^0 + \bm{x}_i$  and expanding $V$ in powers of $\bm{x}_i$. This procedure yields three terms: the fixed position dipolar Hamiltonian (\eq{VddDC}) described in the prior sections, the phonon Hamiltonian which is gapless (see Appendix A) and a phonon-polariton interaction  Hamiltonian $V_{phon-po}$ given by

\begin{equation}
\label{eq_phonon}
\begin{split}
V_{phon-po} &= -2\kappa \sum_{i\neq j}G_{ij}\vec{\sigma}_i\cdot
\vec{\sigma}_j  -2\xi \sum_{i\neq j} G_{{ij}} \sigma^z_i\sigma_j^z\\
&\qquad + 8 B_0 \sum_{i\neq j} G_{{ij}}(\sigma^z_i+\sigma_j^z),
\end{split}
\end{equation}
where
\[
G_{ij} = -3
a^3\frac{(\bf{x}_i-\bf{x}_{j})\cdot(\bf{r}_i^0-\bf{r}_{j}^0)}{4|\bf{r}_i^0-\bf{r}_{j}^0|^5}. \nonumber
\] and $B_0=(\mu_{ee}^2 - \mu_{gg}^2)/( 8 \pi  a^3 \epsilon_0)$.
 $V_{phon-po} $ is not spherically symmetric, Dicke states are not eigenstates of it and consequently  $V_{phon-po} $ will induce  transition outside the Dicke manifold even for the one-photon excitation state, $|1\rangle$. This transitions can degrade the phase gate significantly since they are not suppressed by the MPM due to the gapless nature of the phonon spectrum. If at time $t=0$ we prepare our state in the
$J=N/2$ manifold, the projection of the evolving state on the $n-$ Dicke state can
 be written as
\begin{equation}
\ket{\Psi_n(t)} = e^{-\frac{\gamma_{n,ph}(
t)}{2}t} e^{-i\theta_n (t)} \ket{^N\!/_2 ,-^N\!/_2+n}.
\end{equation} and the decoherence rates $\gamma_{1ph}$ and $\gamma_{2ph}$  approximately calculated using perturbation theory \cite{rabl2}. This procedure  yields an expression for $\gamma_{1ph}$  given by:
\begin{eqnarray}
\gamma_{1ph} &\simeq&  \frac{\pi (\xi+4 B_0)^2}{\hbar\sqrt{\beta}}\nonumber\\
& &\sum_\lambda  \int\frac{d^D{(a\bm{k})}}{(2 \pi)^D} g_\lambda({\bm{k}})
[(N(\omega_\lambda(\bm{k}))+1) \delta(\omega_\lambda({\bm{k}})-\omega_{\bm{k}})\nonumber\\
& &+N(\omega_\lambda({\bm{k}}))\delta(\omega_\lambda({\bm{k}}) +\omega_{\bm{k}})].\notag\\
\label{fgr}
\end{eqnarray}Here $N(\omega_{\lambda}(\mathbf{k}))=1/(e^{\hbar \omega_{\lambda}(\mathbf{k})/(k_BT)} - 1)$ is the thermal occupation
number for phonons with the phonon spectrum $\omega_{\lambda}(\mathbf{k})$ (see \fig{phon1} and \fig{phon}). See \eq{g} for the definition of  $g_\lambda({\bm{k}})$.
For two photon excitations it can be shown in a similar way that
\begin{equation}
\gamma_{2ph}\approx 2 \gamma_{1ph}.
\end{equation}
 Detailed derivations are included in App.~B.

\begin{figure}[h]
\centerline {\includegraphics[width=0.8\linewidth]{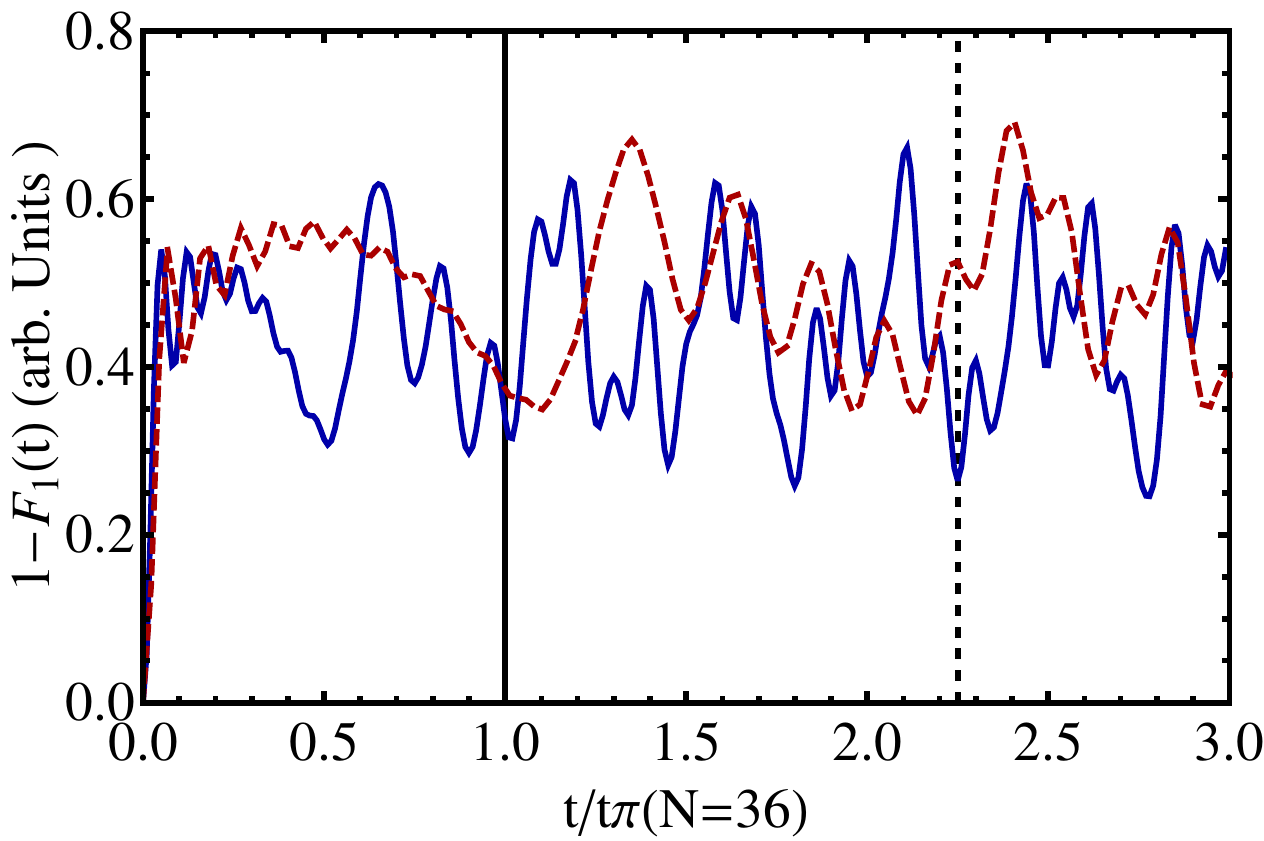}}
\caption{(Color online)$1-F_{1}(t)$, where $1- F_{1}(t)/(\xi+ 4B_0)^2/\sqrt{\beta}$ is the probability of decay of a single excitation
for a 1D system with $N=36$(blue), $N=81$(red-dashed) and  $\xi$, $B_0$ and $\beta$ are system dependant parameters. 
The solid and dashed vertical lines are at the time where the perturbative treatment predicts the implementation of the phase gate for the $N=36$ and $81$ systems respectively.
 }\label{traza1d}
\end{figure}

\begin{figure}[h]
\centerline {\includegraphics[width=0.8\linewidth]{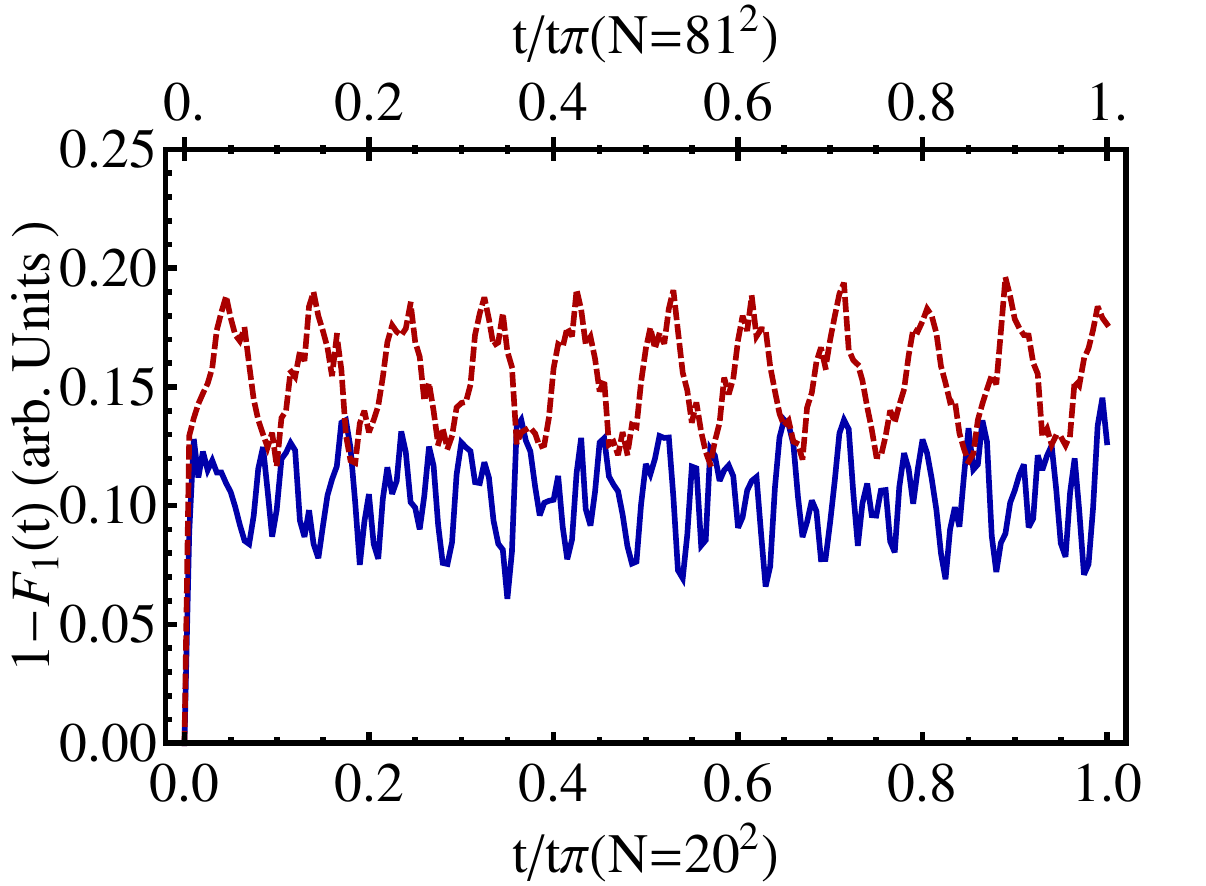}}
\caption{(Color Online) $1-F_{1}(t)$, where $1-F_{1}(t)/(\xi+ 4B_0)^2/\sqrt{\beta}$ is the probability of decay of a single excitation for a 2D system with $N=20^2$(blue), $N=81^2$(red-dashed) and  $\xi$, $B_0$ and $\beta$ are system dependant parameters. In this case we had to go to larger system sizes to show the decrease in fidelity with increasing $N$. The top (bottom) axis scales are for $N=81^2 (20^2)$.
 }\label{traza2d}
\end{figure}

In Ref.~\cite{rabl2} analytical expressions for the decay rates in the thermodynamic limit were derived using the Fermi Golden Rule.  In this limit the decoherence induced by phonons was shown only to be  relevant in 1D and proportional to the temperature.  For our finite number of molecules Fermi Golden rule results are only crude approximations and for a quantitative treatment we instead perform numerical calculations which are summarized in  \fig{traza1d} and \fig{traza2d}. There we plot the time evolution of the decay
probability for two systems with different $N$.
We find  a general tendency of the maximum decay
probability  to grow  with increasing  $N$ specially for 1D.
Also for fixed $N$, $\beta$ and $\kappa$ we observe a linear dependence on the temperature in agreement with   the Fermi Golden Rule predictions.

\section{Conclusion}
In this paper, we explore the feasibility of utilizing cold polar molecules in 1D and 2D optical lattices for coherently controlled nonlinear optics.   We report a controlled $\pi$ phase gate time that increases
proportionally to the number of interacting molecules, but
also note better fidelity in 2D systems for reasonable system
parameters and external field strengths. We address the role of non-symmetric interactions, one of the
 major decoherence effects, and demonstrate the enhancement of phase gate fidelity when an MPM is created by applying external electric fields. For self assembled crystalline samples, we also have explored phonon-induced decoherence.
 Since we find that at low temperature the most relevant decoherence effects in the Wigner crystal arrays are caused by long wave phonon excitations, spin echo techniques could help to reduce them.

\section*{Acknowledgments}
The authors wish to acknowledge NSF for funding, and Peter Rabl for fruitful discussions. Tommaso Calarco wishes to acknowledge the EC Integrated Project SCALA.

\section*{Appendix A}
\label{B}

In a self assembled ensemble molecular crystal, molecules are not longer completely frozen at
the classical equilibrium
positions,  $\bm{r}_i^0$ , which form a linear chain
in 1D or a triangular lattice in 2D, with  lattice spacing $a$. Instead they exhibit collective oscillations (phonons)
in the crystal. If we expand the total Hamiltonian, \eq{Vdd}  around
the equilibrium positions: $ \bf{r}_i=\bf{r}_i^0+\bf{x}_i$ and keep
terms up to quadratic order $\bf{x}_i$  one obtains the following
expression

\begin{figure}[h]
\centerline{\includegraphics[width=1\linewidth]{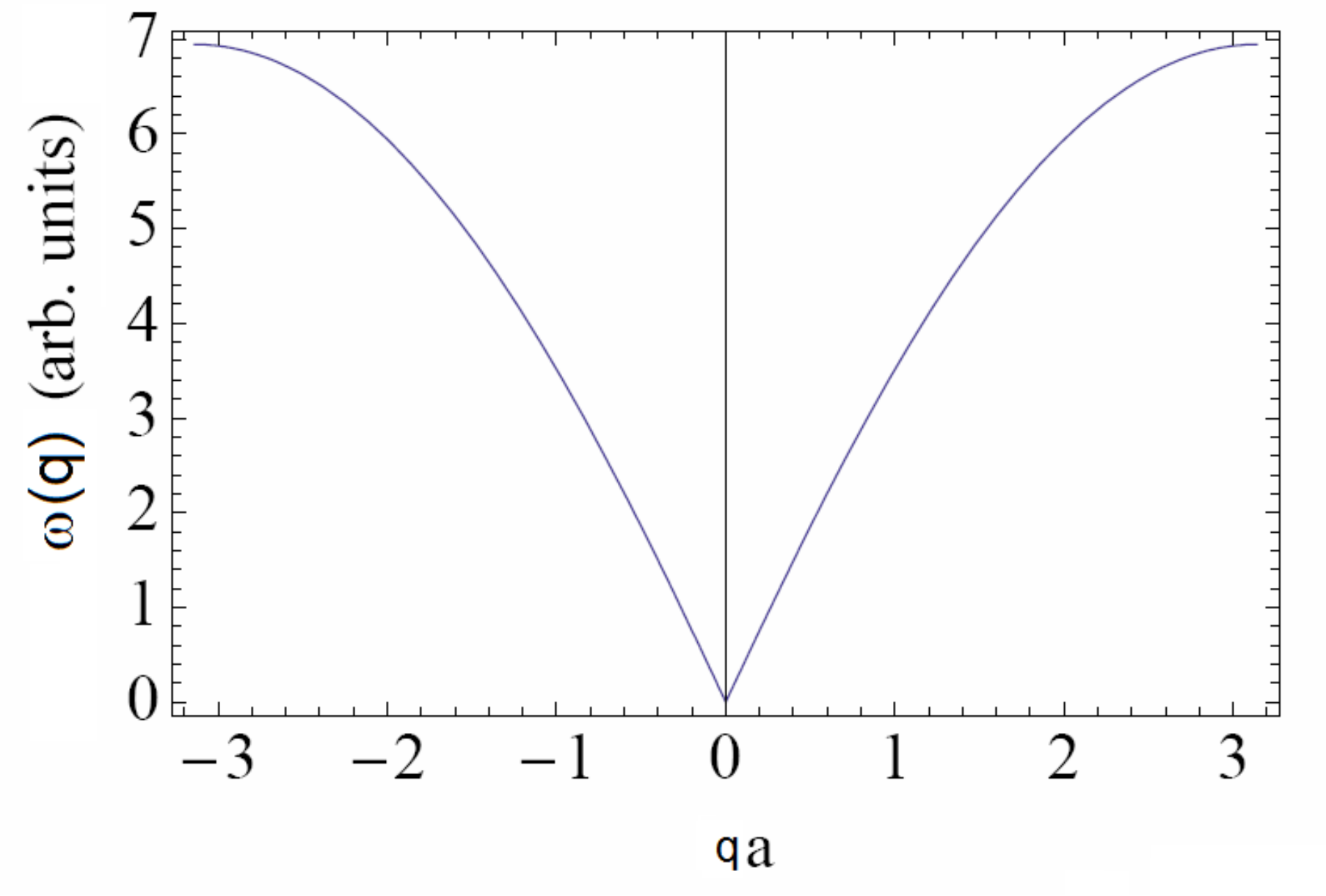}}
\caption{ Phonon excitation spectrum in 1D. $a$ is the separation between the molecules. The units of $\omega(q)$ are $U_{dd}/\sqrt\beta$.}\label{phon1}
\end{figure}

\begin{figure}[h]
\centerline{\includegraphics[width=1\linewidth]{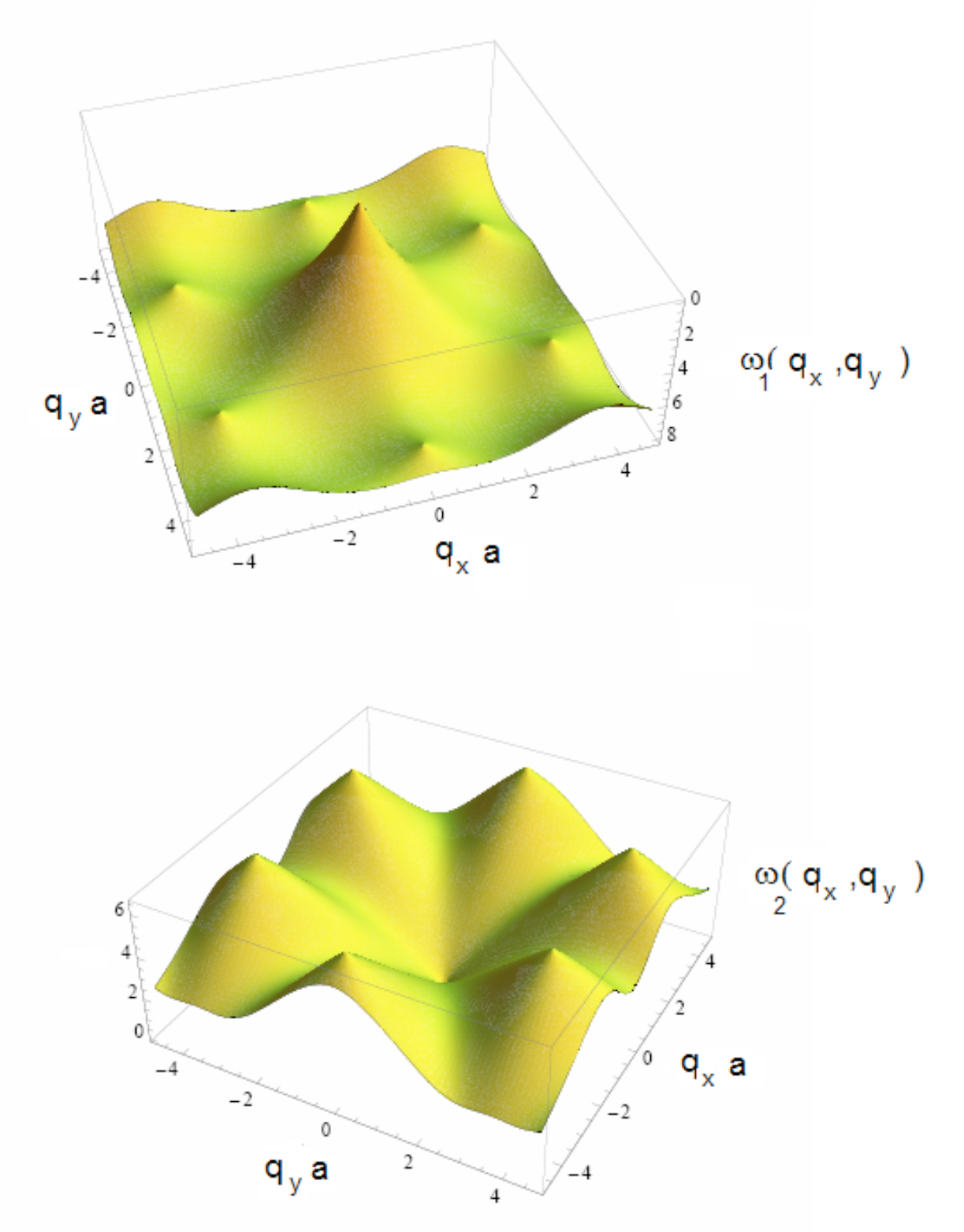}}
\caption{Phonon Excitation spectrum in 2D for the two different branches of the phonon spectrum.  $q_x$ and $q_y$ label the 2D quasimomenta.  The units of $\omega_{1,2} (q_x,q_y)$ are $U_{dd}/\sqrt\beta$.}\label{phon}
\end{figure}

\begin{widetext}

\begin{eqnarray}
H &=& H_{phon}+V+V_{phon-po} \\
H_{phon}&=&\sum_i \frac{\bf{p}_i^2}{2m}+ \frac{3 \mu_{gg}^2}{16\pi \epsilon_0 a^3}\sum_{i\neq
j}\frac{5[(\bf{x}_i-\bf{x}_{j})\cdot
\bf{n}_{ij}^0]^2-[(\bf{x}_i-\bf{x}_{j})]^2}{|\bf{r}_i^0-\bf{r}_{j}^0|^5}\\
&=&\sum_{\bm{q},\lambda} \hbar \omega_\lambda(\bm {q})
\hat{a}^\dagger_{\bm{q},\lambda}\hat{a}_{\bm{q},\lambda}
\end{eqnarray}
\end{widetext}

where $\hat{a}_{\bm{q},\lambda}$ is the annihilation
operator for phonons with quasi-momentum $\bm{q}$ and frequency
$\omega_\lambda(\bm {q})$ and $\bf{n}_{ij}^0$ is a unit vector along $\bf{r}_i^0-\bf{r}_{j}^0 $ . In  2D the index $\lambda$ labels the two
different phonon branches. In general
\begin{eqnarray}
\omega_\lambda(\bm {q})=\frac{U_{dd}}{\sqrt{\beta}}
f_\lambda({\bm{q}} )
\end{eqnarray}
The phonon modes in the dipolar crystal are acoustic phonons,
$f_\lambda({\bm{q}})\sim (c_\lambda \bm{q})$.
The phonon spectrum for 1D and 2D crystals is plotted  in \fig{phon1} and \fig{phon}.

\section*{Appendix B}
\subsubsection{Decay of a single excitation}

We start by deriving the decoherence rate $\gamma_{1ph}$ for
single excitations. To first order in perturbation theory we can write
\begin{equation}
\begin{split}
e^{-\gamma_{1ph}(t)}\simeq
 \qquad 1 -\frac{2}{\hbar^2}\int_{0}^{t}dt'  \int_{0}^{t'}d\tau
\sum_{{\bm {q}},\lambda,\bm {k}>0 }|\mathcal{L}_{{\bm {k}},\bm
{q},\lambda}|^2 \times\\
 \qquad \big[(N(\omega_\lambda(\bm{q}))+1)\cos(\Omega^+_{\bm{q},\bm{k}}
\tau) +N(\omega_\lambda({\bm{q}}))\cos(\Omega^-_{\bm{q},\bm{k}}\tau))\big],
\label{dec}
\end{split}
\end{equation}
where $N(\omega_{\lambda}(\bm{q}))= 1/(e^{\hbar\omega_{\lambda}(\bm{q})/(k_B T)}-1)$ is the
thermal occupation number for phonons with the phonon spectrum  $\omega_{\lambda}(\bm{q})$  (see \fig{phon1} and \fig{phon}) and $\Omega^{\pm}_{\bm{q},\bm{k}}=\omega_\lambda(\bm{q})\pm  \omega_{{\bm k}}$.   $\omega_{{\bm k}}$ is the dispersion relation given in \eq{spinw}. Here we have used the property that

\begin{widetext}
\begin{eqnarray}
{\bf{x}}_i=\frac{1}{\sqrt{N}}\sum_{\bm{q}}\sum_{\lambda=1}^D\sqrt{\frac{
a^2}{2 \sqrt{\beta} f_\lambda(\bm {q})}}
\bm{e}_\lambda\left(\hat{a}_{\lambda,\bm{q}} e ^{i(\bm{q}\cdot \bm
r_i^0-\omega_\lambda(\bm {q}))t}+\hat{a}_{\lambda,\bm{q}}^\dagger e
^{-i(\bm{q}\cdot \bm r_i^0-\omega_\lambda(\bm {q}))t}\right)
\end{eqnarray}.
\end{widetext} where in the 2D case the vectors $\bm{e}_\lambda$  are the two
orthonormal polarization vectors of the two phonon
branches.
$\mathcal{L}_{{\bm {k}},\bm {q},\lambda}= \Bigl| \bra{\psi_0}V_{phon-po}\ket{\psi_k} \Bigr|$, and can be explicitly written as
\begin{eqnarray}
\lefteqn{ \mathcal{L}_{{\bm {k}},\bm {q},\lambda} \;=\;
-3\sqrt{\frac{1}{2N \sqrt{\beta} f_\lambda(\bm {q}) }}
\sum_{i,j}
\frac{a^4(
\bm{e}_\lambda \cdot
(\bm{r}_i^0-\bm{r}_j^0))}{4 |\bf{r}_i^0-\bf{r}_{j}^0|^5}\cdot} \nonumber\\
&& \Bigl[ e^{i \bm
{q} \cdot \bm {r}^0_i} \, \matrixelem{\psi_{\bm{0}} }{-\xi
\sigma_i^z\sigma_j^z+4 B_0
(\sigma_i^z+\sigma_j^z)}{\psi_{\bm{k}}}
\,  \Bigr] \nonumber\\
\label{fgr3}
\end{eqnarray}
With the substitution of
\begin{equation}
g_{\lambda}(\bm {q})\equiv \frac{9}{f_\lambda(\bm {q})}
(\sum_{i\neq 0} \left[ \sin( \bm {q} \cdot \bm {r}^0_i)\frac{( a^4
\bm{e}_\lambda \cdot \bm{r}_i^0)}{|\bf{r}_i^0|^5}\right])^2\label{g}
\end{equation}
\eq{fgr3} can be simplified to:
\begin{equation}
\mathcal{L}_{{\bm {k}},\bm {q},\lambda} =
i\sqrt{\frac{1 }{2 N \sqrt{\beta} }}(\xi+ 4B_0)
\delta_{\bm{q},-\bm{k}} \sqrt{g_{\lambda}(q)}
\label{dec2}
\end{equation}
In  the regime where the  Fermi Golden Rule is expected to be valid the decay rate is constant; $\gamma(t) = \gamma$
and following a similar procedure described in ref.\cite{rabl} one can show it is given by:
\begin{eqnarray}
\gamma_{1ph} &\simeq&  \frac{\pi (\xi+4 B_0)^2}{\hbar\sqrt{\beta}}\label{fe}\\
& &\sum_\lambda  \int\frac{d^D{(a\bm{k})}}{(2 \pi)^D} g_\lambda({\bm{k}})
[(N(\omega_\lambda(\bm{k}))+1) \delta(\omega_\lambda({\bm{k}})-\omega_{\bm{k}})\nonumber\\
& &+N(\omega_\lambda({\bm{k}}))\delta(\omega_\lambda({\bm{k}}) +\omega_{\bm{k}})].\notag
\label{fgr}
\end{eqnarray}

The resonance condition is  defined as  $(\omega_\lambda({\bm{ q}^\pm_0}) \pm
\omega_{\bm{ q}^\pm_0})=0$ and assuming  that $\bm{ q}_0^\pm=0$ is  the only possible solution,
 the  decay rate  is determined by
the  $\bm{k} \to 0$ limit of the integrant  in \eq{fe} which is given in 1D by:

\begin{eqnarray}
\gamma_{1ph}^{1D}\sim \frac{(\xi +
4 B_0)^2}{4}\sqrt{3\zeta(3)}\sqrt{\beta}  k_B T
\end{eqnarray}
Due to the finite value of $\gamma_{1ph}^{1D}$, and the linear dependence on $N$ of $t_\pi $the probability of remaining on the symmetric manifold decreases exponentially with $N$ and decoherence due to phonons  is
certainly a limiting factor.

On the other hand  in 2D, the decay rate vanishes as

\begin{eqnarray}
\gamma_{1ph}^{2D} \propto   \lim _{q\to 0
}  (\xi + 4 B_0)^2
 k_B T  q \to 0
\end{eqnarray} This conclusion however is only a rough estimation  and our numerical simulations shows that phonons can induce important decoherence effects in 2D even in finite crystals.

\subsubsection{Decay of two dipolar excitations}
In the main body of our paper we have stated that  $\gamma_{2ph}\approx 2 \gamma_{1ph}$. A detailed derivation is given below.

To first order in perturbation theory we can write

\begin{widetext}
\begin{eqnarray}e^{-\gamma_{2ph}(t) /\hbar}&\simeq&1 -\frac{2 }{\hbar^2}  \int_{0}^{t}dt'  \int_{0}^{t'}d\tau  \sum_{{\bm {q}},\lambda,\bm {k,k'}>0 }
|\mathcal{S}_{\bm {k}, \bm {k'},\bm {q},\lambda}|^2 \big [
(N(\omega_\lambda(\bm{q}))+1)\cos(\Omega^+_{\bm{q},\bm{k},\bm{k'}}
\tau)+N(\omega_\lambda({\bm{q}}))\cos(\Omega^-_{\bm{q},\bm{k},\bm{k'}}\tau))\big]
\label{decn}
\end{eqnarray}
 where $\Omega^{\pm}_{\bm{q},\bm{k},\bm{k'}}=\omega_\lambda(\bm{q})\pm
(\omega_{\bm {k}}+\omega_{\bm {k'}})$ and $\mathcal{S}_{\bm {k},\bm {k'},\bm {q},\lambda}$ is rewritten in the same form as
\begin{eqnarray}
\mathcal{S}_{\bm {k},\bm {k'},\bm {q},\lambda}&=& -3\sqrt{\frac{1
}{2N \sqrt{\beta} f_\lambda(\bm {q}) }}\sum_{ij}[ e^{i \bm
{q} \cdot \bm {r}^0_i} \langle \psi_{\bm{0}} |-\xi
\sigma_i^z\sigma_j^z +4B_0 (\sigma_i^z+\sigma_j^z)|\psi_{\bm{k},\bm
{k'}}\rangle\frac{a^4( \bm{e}_\lambda \cdot
(\bm{r}_i^0-\bm{r}_j^0))}{4|\bf{r}_i^0-\bf{r}_{j}^0|^5}]\nonumber \\
&=&3i\sqrt{\frac{1}{2N \sqrt{\beta} f_\lambda(\bm {q}) }} \big (
(\xi+ 4B_0)(\delta_{\bm{q},-\bm{k}}\delta_{\bm{k'},\bm{0}}+\delta_{\bm{q},-\bm{k'}}\delta_{\bm{k},\bm{0}})-\frac{4\xi}{N}
\delta_{\bm{-q},\bm{k+k'}} \big )  \times \sum_{i\neq0}\left[
\sin( \bm {q} \cdot \bm {r}^0_i)\frac{( a^4 \bm{e}_\lambda \cdot
\bm{r}_i^0)}{|\bf{r}_i^0|^5}\right] \label{decn2}
\end{eqnarray}

In the above expression the term proportional to $\xi/N$ is much smaller than the one proportional to $\xi + 4B_0$ and can be neglected. Under this assumption,  $\gamma_{2ph}(t) \to \gamma_{2ph}$,
\begin{equation}
\gamma_{2ph}\approx 2 \gamma_{1ph}
\end{equation}

\end{widetext}

\bibliography{mybib}{}

\end{document}